\def \tr {{\rm tr}\,}
\documentclass[twocolumn,showpacs,preprintnumbers,amsmath,amssymb]{revtex4}
\usepackage{graphicx}

\begin{document}

\title{Efficient implementation of finite volume methods in Numerical Relativity}

\author{Daniela Alic, Carles Bona,  Carles Bona-Casas and Joan Mass\'{o}}

\affiliation{Departament de Fisica, Universitat de les Illes
Balears, Palma de Mallorca, Spain}

\pacs{04.25.Dm}

\begin{abstract}
Centered finite volume methods are considered in the context of
Numerical Relativity. A specific formulation is presented, in
which third-order space accuracy is reached by using a
piecewise-linear reconstruction. This formulation can be
interpreted as an 'adaptive viscosity' modification of centered
finite difference algorithms. These points are fully confirmed by
1D black-hole simulations. In the 3D case, evidence is found that
the use of a conformal decomposition is a key ingredient for the
robustness of black hole numerical codes.
\end{abstract}
\maketitle


\section{Introduction}
In recent times, many numerical relativity groups have performed
long-term binary-black-hole (BBH) simulations. This was a long
sought goal, with the main objective of computing gravitational
wave patterns that can be used as templates for detection. The BBH
case was specially relevant in this respect because it is assumed
to be the most likely candidate for detection by the current
ground-based interferometric facilities. This can explain why the
focus in these simulations has been placed in the accurate
modelling of the wave zone: the numerical boundaries are placed
safely far away, which implies the use of large computational
domains. Also, the ability to extract the gravitational wave
signal from the wave zone evolution requires the simulation to
last for quite a long time. These facts, together with the use of
some form of mesh refinement in order to ensure the required
accuracy, make BBH simulations very demanding for the
computational point of view, requiring a big computing
infrastructure.

Black hole simulations, however, deserve some interest by
themselves, independently of the quest for gravitational waves.
One can focus for instance on the strong field region, which can
be modelled by using modest-size computational domains. In this
case, one must refrain from excising the black hole interior,
although many interesting results have been obtained by using
excision~\cite{AB01}, even in cases with some matter
content~\cite{Baiotti05}. The consequences of this choice are well
known (see Ref.~\cite{BR06} for a very clear recent example):
\begin{itemize}
    \item A singularity-avoidant gauge condition must be used in
    order to prevent a singularity to form inside the
    computational domain in a finite amount of coordinate time.
    \item This makes the lapse to collapse in the black hole interior
    zones, while keeping its initial profile in the outer region.
    \item As a consequence, steep lapse gradients appear near the
    apparent horizon, which challenge the stability of the
    numerical algorithm.
\end{itemize}

Most of the current BH simulations are performed with finite
difference algorithms. Regarding space accuracy, the most common
approach is to use a centered fourth-order accurate method,
combined with some artificial dissipation term (Kreiss-Oliger
dissipation)~\cite{GKO95}. The leading error in the solution is
precisely the artificial dissipation one, usually of fourth order.
One can interpret this combination just as a particular
third-order scheme with some built-in dissipation, which can be
tuned by a single parameter. This may be a difficulty in some
cases, where dealing with the black hole interior would require an
amount of dissipation which can be instead too big for the
exterior region (see for instance Ref.~\cite{BR06}). Our point is
that centered Finite Volume methods can provide alternative
third-order accurate algorithms in which the built-in dissipation
is automatically adapted to the requirements of either the
interior or exterior black hole regions.

Finite Volume (FV) methods have a reputation of being
computationally expensive, a price that is not worth to pay for
spacetime simulations, where the dynamical fields usually have
smooth profiles. From this point of view, centered FV methods can
provide some improvement, because the they do not require the full
characteristic decomposition of the set of dynamical fields: only
the values of the propagation speeds are needed~\cite{Leveque2}.

This point can be illustrated by comparing the classical FV
techniques implemented in a previous work~\cite{ABMS99} (hereafter
referred as paper I) with the new FV methods presented here. In
paper I, the general relativistic analogous of the Riemann problem
must be solved at every single interface. This implies
transforming back and forth between the primitive variables (the
ones in which the equations are expressed) and the characteristic
ones (the eigenvectors of the characteristic matrix along the
given axis). In the present paper, a simple flux formula is
applied directly on the primitive variables, so that switching to
the characteristic ones is no longer required. The flux formula
requires just the knowledge of the characteristic speeds, not the
full decomposition.

Another important difference is that in paper I, the primitive
quantities where reconstructed from their average values in a
piecewise linear way, using a unique slope at every computational
cell. Only (piecewise) second order accuracy can be achieved in
this way, so that going to (piecewise) third order would require
the use of 'piecewise parabolic methods' (PPM), with the
corresponding computational overload. In this paper instead we
split every flux into two components before the piecewise-linear
reconstruction (flux-splitting approach~\cite{Leveque2}). This
allows using a different slope for every flux component: this
extra degree of freedom allows us to get (piecewise) third order
accuracy for a specific choice of slopes, without using PPM.

It is true that third-order convergence is rarely seen in
practice. In the context of Computational Fluid Dynamics (CFD),
this is due to the arising of physical solutions (containing
shocks or other discontinuities) which are just piecewise smooth.
These discontinuities can propagate across the computational
domain and the convergence rate is downgraded as a result in the
regions swept away by the discontinuity front. A similar situation
is encountered in black hole evolutions. The use of singularity
avoidant slicing conditions produces a collapse in the lapse
function. As it can be seen in Fig.~\ref{run1D}, a steep gradient
surface is formed (the collapse front) that propagates out as the
grid points keep falling into the black hole. We will see that
this results into a downgrade of accuracy in the regions close to
the collapse front.

Stability problems can also arise from the lack of resolution of
the collapse front, which is typically located around the apparent
horizon. The reconstruction procedure can lead there to spurious
oscillations, which introduce high-frequency noise in the
simulation. In paper I, this problem was dealt with the use of
standard slope limiters, which were crucial for the algorithm
stability. In the present paper, although slope limiters are also
discussed for completeness, their use is not even required in any
of the presented simulations. The new algorithm gets rid by itself
of the high-frequency noise, even for the steep (but smooth)
profiles appearing around the black-hole horizon.

With all these simplifications, the proposed centered FV method
can be interpreted just as an 'adaptive viscosity' generalization
of the finite difference (FD) algorithms discussed before.
Moreover, in the FV context, boundary conditions can be imposed in
a simple way by the 'ghost point' technique. This allows one to
avoid the complications related to the corners and edges treatment
that usually appear in the FD context.

The paper is organized as follows: we present in Section II a
brief summary of the simplest FV methods. In Section III, the
flux-splitting variant is considered, and we show how third-order
space accuracy can be obtained by using just linear
reconstruction. The resulting method is then tested for the
one-dimensional (1D) black-hole in Section IV. Long term (up to
1000m) simulations are performed with a single numerical grid of a
limited resolution, showing the efficiency of the algorithm. A
convergence test is also performed, which confirms the predicted
third-order accuracy in the outside region. The three-dimensional
(3D) black-hole case is considered in Section V. A low resolution
simulation is presented, showing the key role of controlling the
trace of the extrinsic curvature in order to avoid numerical
instabilities. This explains the advantage of using $\tr K$ as a
primitive variable, like in the Conformal ADM (CADM)
formalism~\cite{Nakamura87}. This explains also why a conformal
decomposition was also required for obtaining robust 3D
simulations in paper I, even when using FV methods~\cite{ABMS99}.

For the sake of clarity, the more technical points: stability
analysis, time evolution algorithms and the full explicit form of
the equations, are described in Appendices A, B and C,
respectively.

\section{Centered Finite Volume methods: Flux formulae}

Let us consider the well known 3+1 decomposition of Einstein's
field equations. The extrinsic curvature $K_{ij}$ is considered as
an independent dynamical field, so that the evolution system is of
first order in time but second order in space. Let us transform it
into a fully first order system by considering also the first
space derivatives of the metric as independent quantities. This
requires additional evolution equations for these space
derivatives, that can be obtained in the standard way by permuting
space and time derivatives of the metric, that is
\begin{equation}\label{first_derivs}
    \partial_t~(\partial_k~g_{ab}) =
    \partial_k~(\partial_t~g_{ab})~,
\end{equation}
so that the resulting first order system will describe the same
dynamics than the original second order one.

In this first order form, Einstein's field equations can always be
expressed as a system of balance laws~\cite{BM89}. The evolution
system can be written in the form
\begin{equation}\label{balance_law}
        \partial_t ~\textbf{u} + \partial_k ~\textbf{F}^k (\textbf{u})=
        \textbf{S}(\textbf{u}) ~,
\end{equation}
where both the Flux terms $\textbf{F}$ and the Source terms
$\textbf{S}$ depend algebraically on the array of dynamical fields
$\textbf{u}$, which contains the metric and all its first
derivatives. The terms 'Fluxes' and 'Sources' come from the
hydrodynamical analogous of the system (\ref{balance_law}).

The balance law form is well suited for FV discretization methods.
The idea is to evolve the average of the dynamical fields
$\textbf{u}$ on some elementary cells, instead of evolving just
point values like in the FD approach. The space discretization can
be obtained by averaging (\ref{balance_law}) over an elementary
cell and applying the divergence theorem to get:
\begin{equation}\label{balance_integral}
    \partial_t ~\mathbf{\bar{u}}
+ \oint\textbf{F}^k~dS_k = \mathbf{\bar{S}} \,,
\end{equation}
where the overlines stand for space averages. The evaluation of
partial space derivatives has been replaced in this way by that of
surface integrals of the flux terms.

Let us consider for simplicity the one-dimensional case. We can
start from a regular finite difference grid. The elementary cell
can then be chosen as the interval $(x_{i-1/2}\,,~x_{i+1/2})$,
centered on the generic grid point $x_i$. The dynamical fields
$\mathbf{u}$ can be modelled as piecewise linear functions in
every cell (linear reconstruction, see Fig.~\ref{recon}), so that
the average values $\mathbf{\bar{u}}_i$ coincide with the point
values $\mathbf{u}_i$. The corresponding (first-order accurate) FV
discretization of (\ref{balance_integral}) is then given by
\begin{eqnarray}\label{balance_FD}
    \textbf{u}_i^{n+1} = \textbf{u}_i^n &-& \frac{\Delta t}{\Delta
    x}~[~\textbf{F}^x_{i+1/2}-\textbf{F}^x_{i-1/2}~]
    + \Delta t~\textbf{S}_i\,.
\end{eqnarray}
We will restrict ourselves to these linear reconstruction methods
in what follows.

\subsection*{Flux formulae}

The generic algorithm (\ref{balance_FD}) requires some
prescription for the interface fluxes $\textbf{F}^x_{i\pm 1/2}~$.
A straightforward calculation shows that the simple average
\begin{equation}\label{centered}
        F_{i+ 1/2} ~=~ \frac{1}{2}~(F_i + F_{i+1})
\end{equation}
makes (\ref{balance_FD}) fully equivalent to the standard second
order FD approach. As it is well known, this choice is prone to
developing high-frequency noise in presence of steep gradients,
like the ones appearing in black hole simulations. For this
reason, artificial viscosity terms are usually required in order
to suppress the spurious high-frequency modes~\cite{GKO95}.

\begin{figure}
\includegraphics[width=0.5\textwidth,height=5cm]{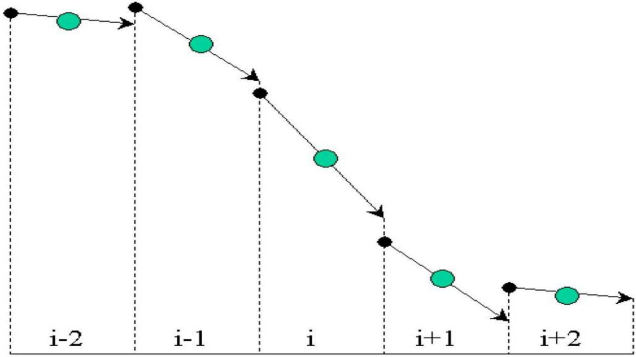}
\caption{Piecewise linear reconstruction of a given function.
Numerical discontinuities appear at every cell interface (dotted
lines) between the left and right values (arrows and dots,
respectively). Note that the original function was monotonically
decreasing: all the slopes are negative. However, both the left
interface values (at $i+3/2$) and the right interface ones (at
$i-3/2$) show local extremes that break the monotonicity of the
original function.} \label{recon}
\end{figure}

We will consider here more general flux formulae, namely
\begin{equation}\label{FLR}
        F_{i+ 1/2} ~=~ f(u_L\,,~ u_R)\,,
\end{equation}
where $u_L$, $u_R$ stand for the left and right predictions for
the dynamical field $u$ at the chosen interface (arrows and dots,
respectively, in Fig.~\ref{recon}). In the (piecewise) linear
case, they are given by
\begin{equation}\label{lrpredictions}
    u^L = u_i + 1/2~\sigma_i~ \Delta x \qquad
    u^R = u_{i+1} - 1/2~\sigma_{i+1}~ \Delta x~,
\end{equation}
where $\sigma_i$ stands for the slope of the chosen field in the
corresponding cell.

A sophisticated choice is provided by the 'shock-capturing'
methods (see Ref.~\cite{Leveque2} for a review). The idea is to
consider the jump at the interface as a physical one (not just a
numerical artifact). The characteristic decomposition of (the
principal part of) the system is then used in order to compute
some physically sound interface Flux. These advanced methods have
been common practice in Computational Fluid Dynamics since
decades. They were adapted to the Numerical Relativity context
fifteen years ago~\cite{BM92}, for dealing with the spherically
symmetric (1D) black-hole case. They are still currently used in
Relativistic Hydrodynamics codes, but their use in 3D black hole
simulations has been limited by the computational cost of
performing the characteristic decomposition of the evolution
system at every single interface.

More recently, much simpler alternatives have been proposed, which
require just the knowledge of the characteristic speeds, not the
full characteristic decomposition. Some of them have yet been
implemented in Relativistic Hydrodynamics codes~\cite{API99}.
Maybe the simplest choice is the local Lax-Friedrichs (LLF) flux
formula~\cite{LLF}
\begin{equation}\label{HLLE}
    f(u_L\,,~ u_R) = \frac{1}{2}~[~F_L + F_R + c~(u_L-u_R )~]~,
\end{equation}
where the coefficient $c~$ depends on the values of the
characteristic speeds at the interface, namely:
\begin{equation}\label{HLLEbc}
    c = max(~\lambda_L\,, \lambda_R~)~,
\end{equation}
where $\lambda$ is the spectral radius (the absolute value of the
biggest characteristic speed).

When comparing the LLF choice (\ref{HLLE}) with the centered FD
one (\ref{centered}), we can see that the supplementary terms play
the role of a numerical dissipation. In this sense, a much more
dissipative choice would be
\begin{equation}\label{LF}
    c = \frac{\Delta x}{\Delta t}~,
\end{equation}
which corresponds to (a piecewise linear generalization of) the
original Lax-Friedrichs algorithm. Note that in any case the
values of the dissipation coefficients are prescribed by the
numerical algorithms: they are no arbitrary parameters, like in
the FD case.

\section{Flux splitting approach}

In the flux formulae approach (\ref{FLR}), the information coming
from both sides is processed at every interface, where different
components are selected from either side in order to build up the
flux there. We will consider here an alternative approach, in
which the information is processed instead at the grid nodes, by
selecting there the components of the flux that will propagate in
either direction (flux splitting approach)~\cite{Leveque2}.

The flux-splitting analogous of the original LLF formula
(\ref{HLLE}, \ref{HLLEbc}) can be obtained by splitting the flux
into two simple components
\begin{equation}\label{pmFlux}
    F_i^{\pm} = F_i \pm \lambda_i~ u_i\,,
\end{equation}
where $\lambda$ will be again the spectral radius at the given
grid point. Each component is then reconstructed separately,
leading to one-sided predictions at the neighbor interfaces. The
final interface flux will be computed then simply as
\begin{equation}\label{pmaverag}
    F_{i+ 1/2} ~=~ \frac{1}{2}~(F^+_L + F^-_R)\,.
\end{equation}
This method can also be expressed as a modified LLF flux formula,
namely
\begin{equation}\label{newFF}
    f(u_L\,,~ u_R) = \frac{1}{2}~[~F_L + F_R
    + \lambda_L~u_L - \lambda_R~u_R ~]~.
\end{equation}

The main difference between the original LLF flux formula
(\ref{HLLE}) and the flux-splitting variant (\ref{newFF}) is that
in the last case there is a clear-cut separation between the
contributions coming from either the left or the right side of the
interface, as it can clearly be seen in (\ref{pmaverag}). In this
way, one has a clear vision of the information flux in the
numerical algorithm. The information from $F^+$ components
propagates in the forward direction, whereas the one from $F^-$
components propagates backwards. This simple splitting provides in
this way some insight that can be useful for setting up suitable
boundary conditions. Moreover, it opens the door to using
different slopes for the reconstruction of each flux component. We
will see below how to take advantage of this fact in order to
improve space accuracy.

\subsection*{Third order accuracy}

As it is well known, the use of a consistent piecewise-linear
reconstruction results generically into a second-order space
accuracy. A convenient choice is given by the centered slope
\begin{equation}\label{fromm}
    \sigma^C = \frac{1}{2\Delta x}~(u_{i+1} - u_{i-1}).
\end{equation}
This is a good default choice (Fromm choice~\cite{Leveque2}),
leading to reliable second-order accurate algorithms .

More general second-order algorithms can be obtained by replacing
the centered slope $\sigma^C$ by any convex average of the left
and right slopes,
\begin{equation}\label{slopes}
    \sigma^L = (u_{i}-u_{i-1})/\Delta x ~,\qquad
    \sigma^R = (u_{i+1}-u_i)/\Delta x~.
\end{equation}
In some applications, however, second order accuracy is not
enough. The leading (third order) error is of the dispersion type,
affecting the numerical propagation speeds. In the FD approach,
this can be improved by using a fourth-order-accurate algorithm in
combination with a fourth-order artificial dissipation term (which
constitutes itself the leading error term). The resulting
combination is third-order accurate.

In the standard FV approach, the standard way of getting
(piecewise) third-order accuracy would be instead to replace the
piecewise linear reconstruction by a piecewise parabolic one. The
prototypical example is provided by the well known piecewise
parabolic methods (PPM). The main complication of this strategy is
that node values would no longer represent the cell averages of a
given dynamical field. This would increase the complexity of the
reconstruction process and the computational cost of the resulting
algorithm.

There is a much simpler alternative, which takes advantage of the
Flux splitting (\ref{pmFlux}). The idea is to consider the
resulting one-sided components $F^\pm$ as independent dynamical
fields, each one with its own slope. The surprising result is that
the choice
\begin{equation}\label{pmsigma}
    \sigma^+ = \frac{1}{3}~\sigma^L + \frac{2}{3}~\sigma^R ~,\qquad
    \sigma^- = \frac{2}{3}~\sigma^L + \frac{1}{3}~\sigma^R
\end{equation}
leads, after the recombination (\ref{pmaverag}), to a third-order
accurate algorithm. The coefficients in (\ref{pmsigma}) are
unique: any other combination leads just to second-order accuracy.

Note that we are getting in this way third-order accuracy with a
piecewise linear reconstruction (see the convergence test in
Fig.~\ref{conv} for a confirmation). This important result seems
to be a peculiarity of the Flux-splitting approach. In order to
better understand it, let us suppress for a moment the lambda
terms in (\ref{pmFlux}-\ref{newFF}). A straightforward calculation
shows that, when using the slopes (\ref{pmsigma}), the resulting
algorithm coincides exactly with the standard
fourth-order-accurate FD algorithm. Adding the lambda terms
improves the stability of the algorithm at the price of
downgrading the space accuracy to third order. This is precisely
the same effect that the Kreiss-Oliger dissipation terms produce
in the FD case. This confirms our result and suggests the
interpretation of the algorithm (\ref{pmFlux}-\ref{newFF}) as
providing an adaptive generalization of the standard dissipation
terms.

\section{The 1D Black Hole}
As a first test, let us consider the Schwarzschild Black Hole in
spherical coordinates. We will write the line element in the
'wormhole' form:
\begin{equation}\label{ds2}
    {\rm d}s^2 = -(~tanh\, \eta~)^2~dt^2
    + 4m^2~(~cosh\, \eta/2~)^4~(~d\eta^2 + d\Omega^2~)~,
\end{equation}
which can be obtained from the isotropic form by the following
coordinate transformation
\begin{equation}\label{eta}
    r = m/2~exp\,(~\eta~)~.
\end{equation}

\begin{figure}
\centering
\includegraphics[width=0.5\textwidth]{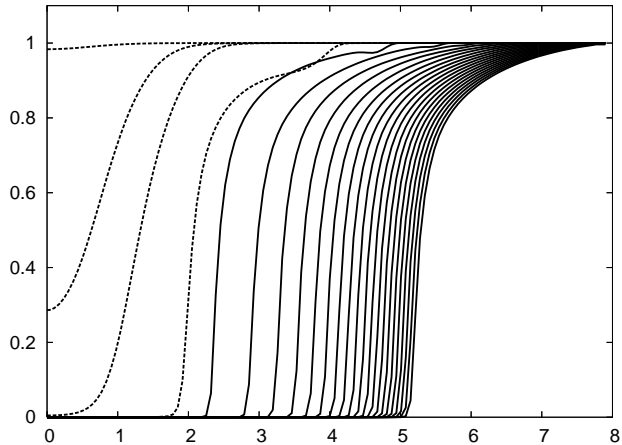}
\caption{Long-term FV simulation of a 1D black hole, with a single
mesh of $120$ gridpoints. The evolution of the lapse is shown up
to $1000m$, in intervals of $50m$ (solid lines). The dotted lines
correspond to $1m$, $3m$, $5m$ and $25m$. Note that the plots tend
to cumulate at the end, due to the exponential character of the
grid, as given by (\ref{eta}). No slope limiters have been used in
this simulation.} \label{run1D}
\end{figure}

The wormhole form (\ref{ds2}) exploits the presence of a minimal
surface (throat) at $\eta=0$. It is manifestly invariant by the
reflection isometry
\begin{equation}\label{isometry}
    \eta \leftrightarrow -\eta~,
\end{equation}
so that the numerical simulations can be restricted to positive
values of $\eta$. The isometry (\ref{isometry}) provides a very
convenient boundary condition at the throat. Moreover (\ref{eta})
implies
\begin{equation}\label{deta}
    dr = r~d\eta\,
\end{equation}
so that an evenly spaced grid in $\eta$ corresponds to a
geometrically increasing spacing in $r$. We can perform in this
way long term simulations with a single grid of a limited size, as
we will see below. This also allows to apply the standard boundary
conditions in FV methods: two 'ghost' points are added by just
copying the nearest neighbor values (or their time variation) for
every dynamical field. The separation between incoming and
outgoing information is automatically performed by the
flux-splitting algorithm, so that boundary points are not special
in this respect.

The simulations are performed with a spherically symmetric version
of the Z3 formalism~\cite{Z3}, as detailed in Appendix C. The free
parameter $n$, governing the coupling with the energy constraint,
is taken with unit value by default, but other similar values can
be taken without affecting significatively the results, like
$n=4/3$, which corresponds to the CADM case~\cite{Z48}. Regarding
gauge conditions, we are using the generalized harmonic
prescription for the lapse~\cite{BM95}
\begin{equation}\label{lapse}
       (\partial_t -{\cal L}_{\beta})\, \alpha
       = - f\;\alpha^2\;tr K
\end{equation}
with zero shift (normal coordinates). We take a constant (unit)
value of the lapse as initial data. We can see in Fig.~\ref{run1D}
the evolution of the lapse in a long-term simulation (up to
$1000m$). We have chosen in this case $f=2/\alpha$ (corresponding
to the 1+log slicing), but similar results can be obtained with
many other combinations of the form
\begin{equation}\label{fchoices}
    f = a + b/\alpha~,
\end{equation}
where $a$ and $b$ are constant parameters.

\begin{figure}
\centering
\includegraphics[width=0.5\textwidth]{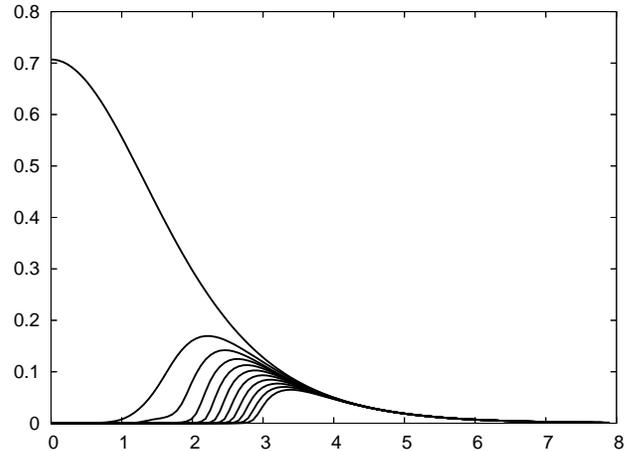}
\caption{The evolution of the propagation speed is shown up to
$100m$, in intervals of $10m$, for the same simulation as in
Fig.~\ref{run1D}. The maximum values are clearly seen to decrease
in time. Note the exponentially decreasing tail, as a result of
the choice of the radial coordinate.} \label{speed1D}
\end{figure}

Note that no slope limiters have been used in the simulation shown
in Fig.~\ref{run1D}. This can seem surprising at the first sight,
but it can be better understood by looking at the propagation
speed profiles shown in Fig.~\ref{speed1D}. The maximum
propagation speed values decrease with time, due to the lapse
collapse in the black hole interior region. This happens because
the initial speed profile is exponentially decreasing with the
chosen radial coordinate. The same decreasing arises for gauge
speed. As a result, the Courant stability condition becomes less
and less restrictive as the simulation proceeds, allowing us to
take bigger timesteps. We have preferred instead to keep the
initial timestep for the sake of accuracy. As far as all
derivative terms get multiplied by $\Delta t$ in the algorithm
(\ref{balance_FD}), this gives us an extra safety factor that
allows us to avoid using slope limiters.

\begin{figure}
\centering
\includegraphics[width=0.5\textwidth]{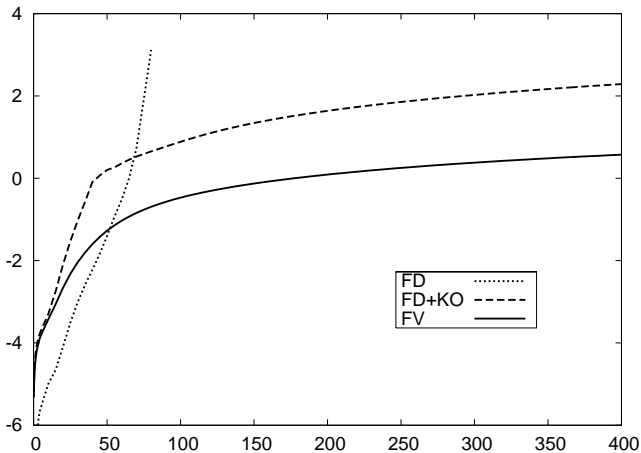}
\caption{Time evolution of the error in the mass function
(logarithm of the $L_2$ norm) for three different numerical
algorithms. The strictly fourth-order FD method, without extra
dissipation terms, is the most accurate as expected, but crashes
after a short time (measured in units of $m$). The other two
algorithms (third-order accurate) get similar errors at early
times, but the FV one performs much better in the long term than
the FD with standard Kreiss-Oliger dissipation. The dissipation
coefficient has been taken as low as allowed by code stability
(see the text). All simulations were obtained with a single mesh
of $120$ gridpoints and using the 1+log slicing prescription. }
\label{mass1D}
\end{figure}

As an accuracy check, we monitor the mass
function~\cite{Lemaitre}, which is to be constant in space and
time for the Schwarzschild case, independently of the coordinate
system. In Fig.~\ref{mass1D}, we compare (the $L_2$ norm of) the
errors in the mass function between a third-order FV simulation
(without slope limiters) and the corresponding FD simulation
(including a fourth order dissipation term like the one in
ref.~\cite{BR06} with $\epsilon = 0.015$). We see that the FD
method shows bigger errors at late times. One can argue that the
leading error in the FD simulation is given by the dissipation
terms, so that one can modify the result by lowering the numerical
dissipation coefficient. However, lowering the viscosity
coefficient used in Fig.~\ref{mass1D}, would result into a
premature code crashing, like the one shown in the Figure for a
strictly fourth order FD run, without the artificial dissipation
term.

We can understand the need for dissipation by looking at the sharp
collapse front in Fig.~\ref{run1D}. We know that this is not a
shock: it could be perfectly resolved by increasing the grid
resolution as needed. In this way we can actually get long-term 1D
black hole simulations, with a lifetime depending on the allowed
resolution. This 'brute force' approach, however, can not be
translated into the 3D case, where a more efficient management of
the computational resources is required. This is where dissipation
comes into play, either the numerical dissipation built in FV
methods or the artificial one which is routinely added to
fourth-order FD methods. Dissipation is very efficient in damping
sharp features, corresponding to high-frequency Fourier modes. As
a result, the collapse front gets smoothed out and can be resolved
without allocating too many grid points. However, the more
dissipation the more error. In this sense, Fig.~\ref{mass1D} shows
that adaptive viscosity built in the proposed FV method provides a
good compromise between accuracy and computational efficiency.

\begin{figure}
\centering
\includegraphics[width=0.5\textwidth]{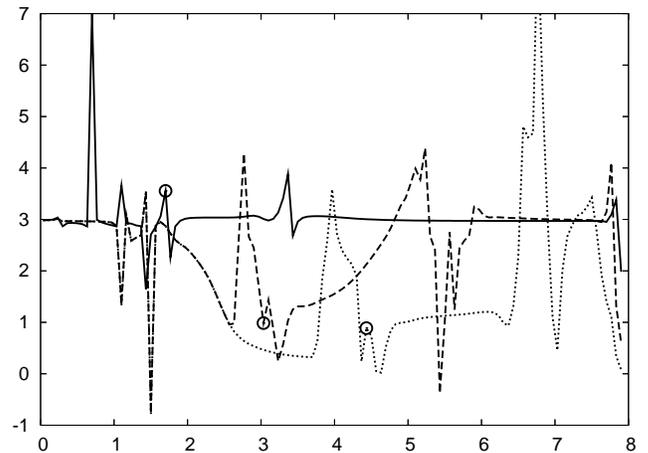}
\caption{Local convergence evolution for the mass function in a 1D
black hole simulation. We can see the predicted third-order
accuracy, when using the proposed slopes (\ref{pmsigma}), around
$t=10\,m$ (solid line). At $t=100\,m$ (dashed line), we yet see
the downgrade in the regions around the collapse front (the
apparent horizon position is marked with a circle). As the
collapse front propagates (dotted line, corresponding to
$t=400\,m$), we can see the growth of the affected regions,
specially the one behind the front.} \label{conv}
\end{figure}

Note that the error comparison is independent of the selected
resolution. This is because the two stable methods in
Fig.~\ref{mass1D} are of third order accuracy, as confirmed by the
local convergence test shown in Fig.~\ref{conv} (solid line,
corresponding to $t=10\,m$). In the long term, however, large
errors develop around the collapse front, downgrading the local
convergence rate in the neighbor regions (dashed and dotted lines
in Fig.~\ref{conv}, corresponding to $t=100\,m$ and $t=400\,m$,
respectively). This can not be seen as a failure of the algorithm
properties, but rather as consequence of large errors in a highly
non-linear context. This also shows that in simulations oriented
to compute gravitational wave patterns (not the case of this
paper), the waveform extraction zone must be safely located, away
both from the outer boundary and from the collapse front.

\section{Preliminary 3D results}

The 1D algorithm (\ref{balance_FD}) can be easily adapted to the
full three-dimensional (3D) case:
\begin{eqnarray}\label{balance_3D}\nonumber
    \textbf{u}_{\{ijk\}}^{n+1} = \textbf{u}_{\{ijk\}}^n
    &-& \frac{\Delta t}{\Delta
    x}~[~\textbf{F}^x_{\{i+1/2\,jk\}}-\textbf{F}^x_{\{i-1/2\,jk\}}~]\\
    \nonumber &-& \frac{\Delta t}{\Delta y}~
    [~\textbf{F}^y_{\{i\,j+1/2\,k\}}-\textbf{F}^y_{\{i\,j-1/2\,k\}}~]\\
    \nonumber &-& \frac{\Delta t}{\Delta
    z}~[~\textbf{F}^z_{\{ij\;k+1/2\}}-\textbf{F}^z_{\{ij\;k-1/2\}}~]\\
    &+& \Delta t~\textbf{S}_{\{ijk\}}\,.
\end{eqnarray}
The structure of (\ref{balance_3D}) suggests dealing with the 3D
problem as a simple superposition of 1D problems along every
single space direction. The stability analysis in Appendix A can
then be extended in a straightforward way, showing that the strong
stability requirement leads to a more restrictive upper bound on
the timestep (in our case, using a cubic grid, this amounts to an
extra $1/3$ factor).

In cartesian-like coordinates, it is not so easy to take advantage
of the reflection isometry (\ref{isometry}). For this reason, we
will evolve both the black-hole exterior and the interior domains.
We can not use the $\eta$ coordinate for this purpose, because the
symmetry center would correspond to $\eta \rightarrow\infty$. We
will take instead the initial space metric in isotropic
coordinates, namely
\begin{equation}\label{isotropic}
    {\rm d}l^2 = (1+\frac{m}{2r})^4~\delta_{ij}~dx^idx^j~.
\end{equation}
We will replace then the vacuum black-hole interior by some
singularity-free matter solution. To be more specific, we will
allow the initial mass to have a radial dependence: $m=m(r)$ in
the interior region. This allows to match a scalar field interior
metric to (\ref{isotropic}) ('stuffed black-hole'
approach~\cite{stuffed}). The price to pay for using a regular
metric inside the horizon is to evolve the matter content during
the simulation: we have chosen the scalar field just for
simplicity.

\begin{figure}
\centering
\includegraphics[width=0.5\textwidth]{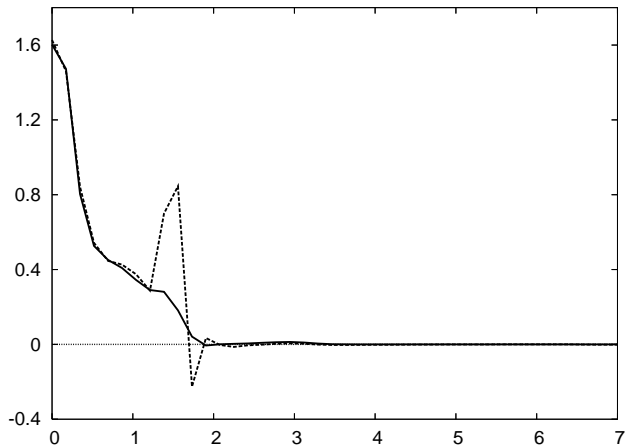}
\caption{Plot of the trace of the extrinsic curvature  at $t=12m$
for a low resolution simulation. The dotted line corresponds to
the trace obtained by contraction from the individual components
$K_{ij}$. The solid line is the same quantity computed directly as
a primitive variable. The big difference corresponds to the
transition between the collapsed and uncollapsed regions, where
the lapse shows a steep profile} \label{trKN}
\end{figure}

We have performed then a low-resolution simulation ($\Delta x =
0.1m$) in order to monitor the errors in $\tr K$, which determines
the evolution of the lapse. We see in Fig.~\ref{trKN} the
comparison between the trace computed by contracting the
individual $K_{ij}$ components (dotted line) and an auxiliary
variable $K$ which is evolved by using the analytical equation for
$\tr K$ (solid line). The difference is striking, even at the
early time of the plot ($t=12m$). Note the negative peak in the
computed $\tr K$, which will produce a spike in the lapse leading
to a premature code crashing.

This behavior could be somehow anticipated from our previous 1D
simulations. The plots shown in Fig.~\ref{run1D} correspond to the
mixed indices equations displayed in Appendix C. We have performed
for comparison the same simulations with 'downstairs' indices and
the results look different. We actually double-checked both codes
before realizing that just raising one index can make a difference
at a given resolution. Of course, in 1D we can always increase
resolution at will and verify that the two results get close
enough. But this would be prohibitive in 3D, at least for
single-grid simulations. Moreover, in 3D we have the additional
difficulty of modelling curved features in a Cartesian grid. In
the spherical case, the worst situation shows up along the main
diagonal, precisely the view shown in Fig.~\ref{trKN}.

These considerations can explain why the CADM formalism, which
actually uses $\tr K$ as a primitive variable, has shown to be
more robust even in single-grid simulations. This also explains
why the use of a conformal decomposition was crucial in the 3D
simulations performed with the old (non-covariant) Bona-Mass\'{o}
formalism in paper I, which used shock-capturing methods. The Z3
formalism can be interpreted as a covariant version of the same,
but our results strongly suggest that the key element for
robustness is not covariance but the use of a conformal
decomposition.

As a final remark, let us focus on the boundary conditions
implementation. The 3D FV algorithm (\ref{balance_3D}) allows to
apply the ghost point technique exactly in the same way as in the
1D case: by just copying (the time variation of) all the
quantities from the neighbor interior point. There is no need for
any special treatment for corners or vertices. Moreover, the
simple FV methods presented here do not require the explicit use
of the characteristic decomposition, not even at the boundaries.
In spite of these simplifications, the robust stability test for
the combined initial-boundary problem gives results equivalent to
the ones obtained with maximally dissipative boundary conditions
in a finite difference context (see Appendix B in
Ref.~\cite{BLPZ05} for details).

{\em Acknowledgements: We acknowledge the hospitality of the
Physics Department at the Louisiana State University during the
final stage of this work. We are specially indebted with Dr.
Palenzuela-Luque fur useful discussions. This work has been
supported by the Spanish Ministry of Science and Education through
the research project number FPA2004-03666 and by the Balearic
Conselleria d'Economia Hissenda i Innovaci\'{o} through the
project PRDIB-2005GC2-06}.

\section*{Appendix A: Stability and Monotonicity}

Let us assume that (the principal part of) the evolution system is
strongly hyperbolic. This means that, for any chosen direction, we
can express the system as a set of simple advection equations for
the characteristic variables (eigenfields). In order to verify the
stability properties of the proposed algorithms, it will be enough
to consider a single advection equation with a generic speed $v$.
The corresponding Flux will be given then by
\begin{equation}\label{advection}
    F(u) = v~u~.
\end{equation}

We will consider in the first place the first-order accurate
approximation, obtained by a piecewise constant reconstruction
(zero slope). The corresponding discretization can be obtained by
replacing the prescription (\ref{pmaverag}) into the general
expression (\ref{balance_FD}). The result is the linear
three-point algorithm:
\begin{eqnarray}\label{avdectionFD}\nonumber
    u_i^{n+1} = u_i^n &+& \frac{\Delta t}{\Delta x}
    ~[~\frac{1}{2}~(\lambda_{i+1}-v_{i+1})~u_{i+1}^n\\
    &+& \frac{1}{2}~(\lambda_{i-1}+v_{i-1})~u_{i-1}^n
    - \lambda_i~u_i^n~]~\,.
\end{eqnarray}
Allowing for the fact that $\lambda$ is chosen at every point as
the absolute value of the maximum speed, we can see that all the
$u^n$ coefficients are positive provided that the Courant
stability condition
\begin{equation}\label{Courant1}
    \lambda~\frac{\Delta t}{\Delta x} \le 1
\end{equation}
is satisfied. Note however that a more restrictive condition is
obtained in the three-dimensional case, where we must add up in
(\ref{avdectionFD}) the contributions from every space direction.

As it is well known, the positivity of all the coefficients
ensures that the algorithm is monotonicity-preserving, so that
spurious numerical oscillations can not appear. This implies
stability, but the converse is not true, as it is well known. Let
us remember at this point that the centered FD discretization
could be recovered from (\ref{avdectionFD}) simply by setting
$\lambda$ to zero, although we would lose the monotonicity
property in this way.

The monotonicity properties of the piecewise constant
reconstruction are not ensured in the piecewise linear case. We
can clearly see in Fig.~\ref{recon} that monotonicity problems can
arise in steep gradient regions. The reason is that either the
series of left $\{u^L\}$ or right $\{u^R\}$ interface predictions
can show spurious peaks which where not present in the original
function. In the case of the centered slope (\ref{centered}), a
detailed analysis shows that this will happen at a given interface
only if the left and right slopes differ by a factor of three or
more. This gives a more precise sense to the 'steep gradient'
notion in the centered slopes case.

The natural way to remedy this is to enforce that both (left and
right) interface predictions are in the interval limited by the
corresponding left and right point values (interwinding
requirement). This amounts to using the 'limited' slopes
\begin{equation}\label{MMC}
    \sigma^{lim} =
    minmod(~2\sigma^L\,,~\sigma\,,~2\sigma^R~)~,
\end{equation}
where $\sigma$ is the default slope at the given cell. This
interwinding requirement is not enough, however, to ensure the
positivity of all the coefficients in the resulting algorithm. A
detailed analysis shows that an extra factor in the Courant
condition would be required for monotonicity in this case:
\begin{equation}\label{Courant2}
    \lambda~\frac{\Delta t}{\Delta x} \le 1/2~.
\end{equation}
Note however that we are analyzing here the elementary step
(\ref{balance_FD}). This is just the building block of the time
evolution algorithm. The exact stability and monotonicity limits
for the time step would depend on the specific choice of the full
time evolution algorithm~\cite{GKO95}, which will be described in
Appendix B.

A word of caution must be given at this point. It is well known
that the monotonicity results hold only for strictly
Flux-conservative algorithms. This is not our case: the Source
terms play an important physical role. Of course, these terms do
not belong to the principal part, so that positivity of the Flux
terms ensures some strong form of stability. Nevertheless, one
must be very careful with the physical interpretation, because the
first-order constraints (\ref{first_derivs}) preclude any
clear-cut isolation of the Source terms. This makes the analogy
with Fluid Dynamics only approximative and the use of the slope
limiters a risky matter: we could be removing in the Flux part
some features that are required to compensate something in the
Source part. Our experience is that, at least for smooth profiles,
more robust numerical simulations are obtained when the slope
limiters are switched off. The high frequency modes are kept under
control by the numerical dissipation built in the proposed FV
methods.

\section*{Appendix B: Time accuracy}
The simple step (\ref{balance_FD}) is
only first-order accurate in time, and this fact is not changed by
any of the space accuracy improvements we have considered up to
now. The standard way of improving time accuracy is by the method
of lines (MoL, see refs.~\cite{MoL}~\cite{GKO95}). The idea is to
consider (\ref{balance_FD}) as a basic evolution step
\begin{equation}\label{step}
    E(~u^n\,,~\Delta t~)
\end{equation}
in order to build higher order algorithms. A convenient choice for
these time evolution algorithms is provided the standard
Runge-Kutta methods~\cite{GST01} (see also ~\cite{GKO95}). For
instance, second order accuracy can be obtained in two steps:
\begin{equation}\label{RK2}
    u^* = E(~u^n,\,\Delta t~)\qquad
    u^{n+1} = \frac{1}{2}~[~u^n + E(~u^*,\,\Delta t~)~],
\end{equation}
and third-order time accuracy with one more intermediate step:
\begin{eqnarray}\label{RK3}\nonumber
    u^{**} &=& \frac{3}{4}~u^n + \frac{1}{4}~E(~u^*\,,~\Delta t~)\\
    u^{n+1} &=& \frac{1}{3}~u^n + \frac{2}{3}~E(~u^{**}\,,~\Delta t~)~.
\end{eqnarray}

Note that the positivity of all the coefficients in (\ref{RK2},
\ref{RK3}) ensures that the monotonicity property of the basic
step (\ref{step}) will be preserved by the resulting
strong-stability-preserving (SSP) algorithm. This interesting
property comes at the price of keeping the upper limit on $\Delta
t$ that is required for the monotonicity of the basic step. This
is a clear disadvantage with respect to the case in which the
standard FD approach is being used for space discretization, in
which one is only limited by plain stability, not monotonicity.
Then, there are Runge-Kutta algorithms (with non-positive
coefficients) that alow to take $\Delta t$ larger than the one
required by the standard Courant condition~\cite{GKO95}.

Conversely, second order Runge-Kutta algorithms like (\ref{RK2})
are unstable when used in combination with FD space
discretization, unless artificial dissipation is added in order to
recover stability (not just monotonicity)~\cite{GKO95}. This is
why FD simulations currently use at least a third-order time
evolution algorithm.

\section*{Appendix C: Z3 evolution equations}

The Z3 evolution system~\cite{Z3,Z48} is given by:
\begin{eqnarray}
  (\partial_t &-&{\cal L}_{\beta})\; \gamma_{ij}
  = - {2\,\alpha}\,K_{ij}
\label{evolve_metric} \\
\nonumber
   (\partial_t &-&{\cal L}_{\beta}) K_{ij} = -\nabla_i\,\alpha_j
    + \alpha\,   [\,R_{ij}
    + \nabla_i Z_j + \nabla_j Z_i
\nonumber \\
     &-& 2\,K^2_{ij}+tr K\,K_{ij}
 - S_{ij}+\frac{1}{2}\,(\,tr S + (n-1)\, \tau\,)\,\gamma_{ij}\,]
\nonumber \\
  &-&\frac{n}{4} ~ \alpha\, [\,tr\,R + 2\; \nabla_kZ^k \nonumber\\
 & &\qquad +4~ tr^2 K - tr(K^2) - 2\,Z^k\alpha_k/\alpha\,]\;\gamma_{ij}
\label{evolve_K} \\
\label{evolve_Zs}
 (\partial_t &-&{\cal L}_{\beta}) Z_i = \alpha\; [\nabla_j\,({K_i}^j
  -{\delta_i}^j~ tr K) -2 {K_i}^j Z_j - S_i]~,\nonumber\\
  & &
\end{eqnarray}
where $n$ is an arbitrary parameter governing the coupling of the
energy constraint.

The fully first-order version can be obtained in the standard way,
by introducing the additional fields
\begin{equation}\label{Dkij}
    D_{kij} \equiv \frac{1}{2}~\partial_k~\gamma_{ij}~.
\end{equation}
Note that the ordering constraint (\ref{first_derivs}) reads
\begin{equation}\label{ordering}
    \partial_r~D_{kij} = \partial_k~D_{rij}~,
\end{equation}
which is no longer an identity for the first order system. As a
consequence of this ordering ambiguity of second derivatives, the
Ricci tensor term in (the first order version of) the evolution
equation (\ref{evolve_K}) can be written in many different ways.
Then, an ordering parameter $\zeta$ can be introduced~\cite{Z48},
so that the parameter choice $\zeta = +1$ corresponds to the
standard Ricci decomposition
\begin{equation}\label{Def3R}
{}^{(3)}\!R_{ij}~=~\partial_k\;{\Gamma^k}_{ij}-\partial_i\;{\Gamma^k}_{kj}
+{\Gamma^r}_{rk}{\Gamma^k}_{ij}-{\Gamma^k}_{ri}{\Gamma^r}_{kj}
\end{equation}
whereas the opposite choice $\zeta = -1$ corresponds instead to
the decomposition
\begin{eqnarray}\label{Def3dDF}
{}^{(3)}\!R_{ij}&=&-\partial_k\;{D^k}_{ij}+\partial_{(i}\;{\Gamma_{j)k}}^{k}
- 2 {D_r}^{rk} D_{kij} \nonumber \\
&+& 4 {D^{rs}}_i D_{rsj} - {\Gamma_{irs}}
{\Gamma_j}^{rs}-{\Gamma_{rij}} {\Gamma^{rk}}_k~,
\end{eqnarray}
which is most commonly used in Numerical Relativity codes. We can
then consider the generic case as a linear combination of
(\ref{Def3R}) and (\ref{Def3dDF}).

In the spherically symmetric vacuum case, the first order version
of the system (\ref{evolve_metric}-\ref{evolve_K}) is free of any
ordering ambiguity. It can be written as
\begin{eqnarray}
    \partial_{t}\, \gamma_{rr} &=&
    -2\,\alpha\,\gamma_{rr}\,K^{~r}_{r}, \qquad
    \partial_{t}\, \gamma_{\theta\theta} =
    -2\,\alpha\,\gamma_{\theta\theta}\,K^{~\theta}_{\theta}
\label{evolve_g} \\
    \partial_{t}\, K^{~r}_{r}
    &+& \partial_{r}[\alpha\,\gamma^{rr}\,(A_{r}+(2-n)\,D_{\theta}^{~\theta}-(2-n/2)\,Z_{r})] =
    \nonumber \\
    & &\alpha\,[(K^{~r}_{r})^{2}+(2-n)\,K^{~r}_{r}\,K^{~\theta}_{\theta}-(n/2)\,(K^{~\theta}_{\theta})^{2}
    \nonumber \\
    &
    &-\gamma^{rr}\,D_{r}^{~r}\,(A_{r}+(2-n)\,D_{\theta}^{~\theta}+(n/2-2)\,Z_{r})
    \nonumber \\
    & & +\gamma^{rr}\,D_{\theta}^{~\theta}\,((2-n)\,A_{r}-(2-3\,n/2)\,D_{\theta}^{~\theta}-n\,Z_{r})
    \nonumber \\
    & & - \,\gamma^{rr}\,(2-n)\,A_{r}\,Z_{r}-(n/2)\,\gamma^{\theta\theta}]
\label{evolve_kr} \\
    \partial_{t}\, K^{~\theta}_{\theta}
    &+& \partial_{r}[\alpha\,\gamma^{rr}\,((1-n)\,D_{\theta}^{~\theta}+(n/2)\,Z_{r})] =
    \nonumber \\
    & &\alpha\,[(1-n)\,K_{r}^{~r}K_{\theta}^{~\theta}\,+(2-n/2)\,(K_{\theta}^{~\theta})^2
    \nonumber \\
    & & -\gamma^{rr}\,D_{r}^{~r}\,((1-n)\,D_{\theta}^{~\theta}+(n/2)\,Z_{r})
    \nonumber \\
    & & +\gamma^{rr}\,D_{\theta}^{~\theta}\,((2-n)\,Z_{r}-(2-3n/2)\,D_{\theta}^{~\theta})
    \nonumber \\
    & & - n\,\gamma^{rr}\,A_{r}\,(D_{\theta}^{~\theta}-Z_{r})
    + (1-n/2)\,\gamma^{\theta\theta}]
\label{evolve_kt} \\
    \partial_{t}\, Z_{r}
    &+& \partial_{r}[2\,\alpha\,K_{\theta}^{~\theta}] =\nonumber\\
    & & 2\,\alpha\,[D_{\theta}^{~\theta}\,(K^{~r}_{r}-K^{~\theta}_{\theta})+A_{r}\,K^{~\theta}_{\theta}-K^{~r}_{r}\,Z_{r}]
\label{evolve_z} \\
    \partial_{t}\, D_{r}^{~r}
    &+& \partial_{r}[\alpha\,K_{r}^{~r}] = 0,
\qquad
    \partial_{t}\, D_{\theta}^{~\theta}
    + \partial_{r}[\alpha\,K_{\theta}^{~\theta}] = 0,
\label{evolve_D}
\end{eqnarray}
where we are using normal coordinates (zero shift). The slicing
condition (\ref{lapse}) can be written as
\begin{equation}\label{evolve_alpha}
    \partial_{t}\, \alpha = -\alpha^{2}\,f\,trK~, \qquad
    \partial_{t}\, A_{r} + \partial_{r}[\alpha\,f\,trK] = 0~.
\end{equation}

The mass function can be defined for spherically symmetric
spacetimes as~\cite{Lemaitre}
\begin{equation}\label{Bondi4D}
    2M = Y~[~1-g^{ab}\partial_a\,Y\,\partial_b\,Y~]~,
\end{equation}
where $Y$ stands for the area radius. In spherical coordinates we
get
\begin{equation}\label{Bondi}
    2M(t,r) = \sqrt{\gamma_{\theta\theta}}~\{~1+
    \gamma_{\theta\theta}\,[(K_{\theta}^{~\theta})^2
    -\gamma^{rr}(D_{\theta}^{~\theta})^2]~\}~.
\end{equation}

The mass function has a clear physical interpretation: it provides
the mass inside a sphere of radius $r$ at the given time $t$. It
follows that $M(t,r)$ must be constant for the Schwarzschild
spacetime, no matter which coordinates are being used. This
provides a convenient accuracy check for numerical simulations.

\bibliographystyle{prsty}

\end{document}